\documentclass[11pt,a4paper]{article}

\usepackage[margin=0.88in]{geometry}
\usepackage[utf8]{inputenc}
\usepackage[T1]{fontenc}
\usepackage{lmodern}
\usepackage{microtype}
\usepackage{setspace}
\usepackage{titlesec}
\usepackage{enumitem}
\usepackage{booktabs}
\usepackage{longtable}
\usepackage{array}
\usepackage{tabularx}
\usepackage{xcolor}
\usepackage{hyperref}
\usepackage{xurl}
\usepackage{fancyhdr}
\usepackage{amsmath,amssymb}
\usepackage{graphicx}
\usepackage{caption}
\usepackage{float}
\usepackage{placeins}
\usepackage{ragged2e}

\definecolor{msnavy}{HTML}{0B1F5B}
\definecolor{msblue}{HTML}{1656C1}
\definecolor{mslight}{HTML}{EEF4FF}
\definecolor{msgreen}{HTML}{1D7A39}
\definecolor{msorange}{HTML}{C85A0A}
\definecolor{msgray}{HTML}{586174}
\definecolor{msrule}{HTML}{C7D1E3}

\hypersetup{
    colorlinks=true,
    linkcolor=msblue,
    urlcolor=msblue,
    citecolor=msblue,
    pdftitle={Mindspeller Neuroprofile White Paper - September 2026},
    pdfauthor={Mindspeller Research and Product Team}
}

\setlist[itemize]{topsep=3pt,itemsep=2pt,parsep=0pt,leftmargin=1.45em}
\setlist[enumerate]{topsep=3pt,itemsep=2pt,parsep=0pt,leftmargin=1.7em}
\titleformat{\section}{\Large\bfseries\color{msnavy}}{\thesection}{0.75em}{}
\titleformat{\subsection}{\large\bfseries\color{msblue}}{\thesubsection}{0.7em}{}
\titleformat{\subsubsection}{\normalsize\bfseries\color{msnavy}}{\thesubsubsection}{0.6em}{}
\titlespacing*{\section}{0pt}{14pt}{6pt}
\titlespacing*{\subsection}{0pt}{10pt}{4pt}
\titlespacing*{\subsubsection}{0pt}{8pt}{3pt}

\newcommand{\mindspeller}{Mindspeller}
\newcommand{\neuroprofile}{Neuroprofile}
\newcommand{\onet}{O*NET}
\newcommand{\callout}[2]{%
  \begin{center}
  \fcolorbox{msrule}{mslight}{%
    \parbox{0.94\textwidth}{\textbf{\color{msnavy}#1}\par\smallskip #2}%
  }
  \end{center}
}
\newcommand{\pipelinearrow}{\ensuremath{\rightarrow}}

\title{\textbf{\color{msnavy}Mindspeller Neuroprofiling}\\[4pt]
\large How task performance, EEG, and association evidence support O*NET-based role guidance}
\author{
Prem Aravindan Jeyakumar \\
Marc M. Van Hulle \\
Hannes De Wachter
}
\date{Updated: September 2026}

\begin{document}

\maketitle
\vspace{-8pt}
\begin{center}
\IfFileExists{logo-with-text.png}{\includegraphics[width=0.56\textwidth]{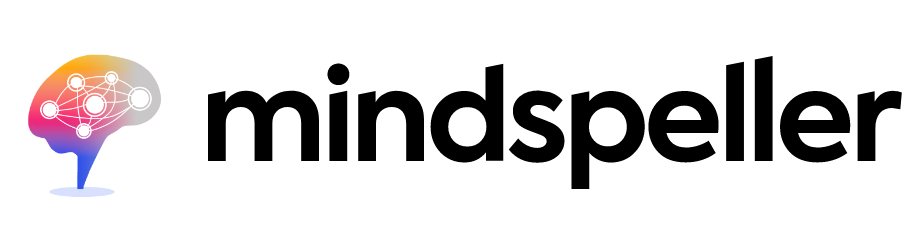}}{}
\end{center}

\callout{Terminology}{\mindspeller{} is the company and platform. A \neuroprofile{} is the final report. Of the three evidence streams used in that report, only the task-and-EEG stream is neurophysiological. Self-report and association evidence help explain the result, but they do not create \onet{} abilities or role candidates.}

\begin{abstract}
\mindspeller{} produces a \neuroprofile{} from three sources: rational self-report, association-based semantic positioning, and performance recorded during cognitive tasks together with EEG. The task-and-EEG stream is the only source used to create occupational evidence. A result can enter role matching only after the participant's task performance supports the intended construct and the corresponding EEG data pass the required quality and evidence checks.

The current pilot uses four EEG electrodes, twelve scored tasks, 23 \onet{} abilities, and an internal bank of 376 occupations. Self-report and association evidence help explain motivation, preference, and alignment, but do not generate roles. The output is intended to support discussion about cognitive fit. It is not a hiring decision, a measure of practical job skill, or a prediction of job performance. Role confidence is currently capped at Moderate, and external psychometric and job-outcome validity have not yet been established.
\end{abstract}

\newpage
\setcounter{tocdepth}{1}
\tableofcontents
\newpage

% -----------------------------------------------------------------------------
\section{Executive Summary}

Recruitment now often takes place through remote interviews, online assessments, automated screening, and AI-assisted workflows. These methods are useful, but each gives only a partial view of a candidate. Resumes and interviews depend heavily on self-presentation. Questionnaires depend on what people know and are willing to say about themselves. Short assessments may miss how a person performs when several cognitive demands are present at once.

The \mindspeller{} \neuroprofile{} brings together three evidence streams:

\begin{itemize}
    \item \textbf{Rational self-report}: the participant's stated preferences, values, goals, and aspirations.
    \item \textbf{Association and archetype evidence}: responses to images, brands, descriptors, concepts, and IAT-style prompts, used to place the participant within the \mindspeller{} semantic network.
    \item \textbf{Task-and-EEG evidence}: scored task performance and EEG changes recorded during a structured cognitive battery.
\end{itemize}

For the final narrative, \mindspeller{} assigns approximately 5\% of the emphasis to self-report, 35\% to association evidence, and 60\% to task-and-EEG evidence. These are product settings chosen to give the greatest emphasis to observed task performance and physiology. They are not psychometric reliability coefficients, and they do not decide whether an ability or role is accepted. Only the task-and-EEG stream can produce occupational evidence.

The role process follows a fixed order:

\begin{center}
\textbf{task performance + matched EEG evidence}
\pipelinearrow{} task-supported characteristic
\pipelinearrow{} allowed \onet{} ability
\pipelinearrow{} role match
\pipelinearrow{} work-style summary
\pipelinearrow{} \neuroprofile{}.
\end{center}

In plain language, a task does not prove an ability simply because it was designed to test that ability. The participant must also produce a valid behavioural result, and the EEG recorded during that task must be usable and consistent with the evidence rules.

The current ability set contains 23 \onet{} abilities. The internal role bank contains 376 occupations: 310 professional or knowledge-work roles and 66 trades or field-operations roles. The user-facing report shows only a small portfolio. Professional roles form the majority. A limited number of trades or field roles may appear as supplementary pathways.

A supplementary pathway is not a recommendation to enter that occupation. It is an example of a work setting that uses a similar cognitive pattern. For example, an Electrician or technical-maintenance pathway may illustrate structured troubleshooting, sequencing, or monitoring. It does not show that the participant has trade skill, physical capacity, sensory acuity, safety readiness, or a licence.

The protocol is still in pilot status. The software includes signal checks, behavioural rules, automated regression tests, and end-to-end verification. These checks show that the implemented process runs as intended. They do not yet show that the \neuroprofile{} predicts training success, job performance, retention, or hiring outcomes. For this reason, role confidence is currently limited to Moderate.

% -----------------------------------------------------------------------------
\section{Problem Context and Design Objective}

\subsection{The Say--Do Discrepancy}

People do not always know how they will process information under demand. A candidate may describe themselves as fast, creative, collaborative, analytical, or detail-oriented, yet perform in a more specific way during structured tasks. They may sequence information well but select responses slowly, maintain focus but find interruptions costly, or generate ideas more effectively when the problem has clear limits.

The \neuroprofile{} keeps these differences visible. Agreement between streams is reported as alignment. Disagreement may reflect aspiration, context, or a genuine difference between self-perception and measured task behaviour. It is not automatically treated as error.

\subsection{Remote Recruitment and Incomplete Evidence}

Remote recruitment gives employers less direct behavioural context. A short interview rarely shows how a candidate updates working memory, detects an anomaly, changes rules, follows a complex instruction, or balances speed against accuracy. \mindspeller{} does not replace interviews, qualifications, work samples, or human judgement. It adds a structured source of evidence that can guide follow-up questions and make the reasoning behind a role discussion clearer.

\subsection{AI as a Pair Worker}

AI now supports many tasks in writing, coding, design, analysis, planning, debugging, and research. The relevant question is therefore not only what a person can do alone, but also how they frame a problem, check an output, switch between demands, and decide when an AI response needs correction. The \neuroprofile{} does not measure AI productivity. It can, however, describe task patterns that may matter in AI-assisted work, such as structured checking, anomaly detection, comprehension, switching, and idea generation.

\subsection{Cognitive-Fit Matching Is Not Hiring Suitability}

Role matching in \mindspeller{} means comparing measured cognitive abilities with the cognitive requirements listed for an occupation. Hiring suitability is a much broader judgement. It also depends on experience, qualifications, domain knowledge, work samples, communication, references, legal eligibility, compensation, availability, and the employer's needs.

A person may show cognitive fit for a role and still lack the experience or qualifications needed for that job. The reverse is also possible: an experienced candidate may be suitable to hire even when the \neuroprofile{} can state only Moderate confidence. The report is intended to inform recruitment discussion, not to make a hire/no-hire decision.

\subsection{Design Objective}

The objective is to connect each occupational statement to the evidence that supports it. The system also separates measured findings from context, caveats, and information that remains outside scope. The final AI-generated text must stay within those limits.

% -----------------------------------------------------------------------------
\section{Neuroprofile Architecture}
\label{sec:architecture}

\subsection{Evidence Streams and Processing Layers}

This paper uses \textbf{stream} for a source of evidence and \textbf{layer} for a processing step. Self-report, association responses, and task-and-EEG data are parallel streams. Feature extraction, interpretation, ability mapping, role matching, work-style summarization, and final synthesis are sequential layers.

\begin{figure}[H]
    \centering
    \IfFileExists{explainable_neuroprofile.png}{%
      \includegraphics[width=\textwidth]{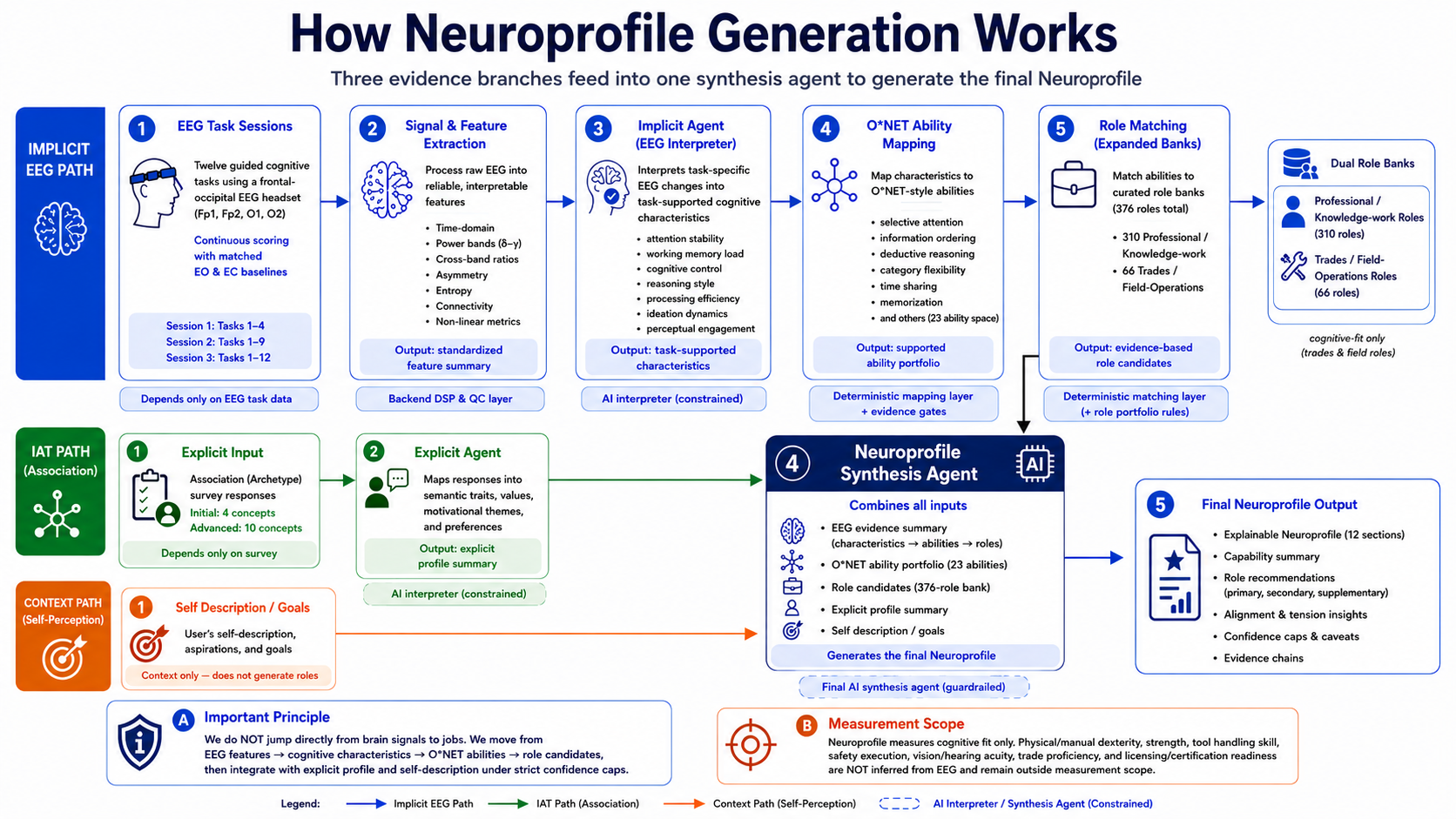}%
    }{%
      \fbox{\parbox{0.94\textwidth}{\centering High-level Neuroprofile architecture figure placeholder.}}%
    }
    \caption{How the three evidence streams move through the \neuroprofile{} system. The blue task-and-EEG branch creates occupational evidence. The green association branch and orange self-report branch add context before the final report is written.}
    \label{fig:neuroprofile_generation}
\end{figure}

The upper branch in Figure~\ref{fig:neuroprofile_generation} is the only route to \onet{} abilities and roles. The other two branches help explain preferences, goals, and agreement or tension between what the participant reports and what the tasks show. All three meet in the final synthesis layer.

\subsection{Current Measurement and Evidence Architecture}

Table~\ref{tab:current_architecture} gives a compact overview of the current implementation. Detailed processing rules remain in the internal technical documents \cite{internalredesign}.

\begin{table}[H]
\centering
\small
\begin{tabularx}{\textwidth}{>{\RaggedRight\arraybackslash}p{0.22\textwidth}>{\RaggedRight\arraybackslash}p{0.39\textwidth}X}
\toprule
\textbf{Component} & \textbf{Current implementation} & \textbf{Why it is included} \\
\midrule
EEG recording configuration & Four electrodes: Fp1/Fp2 over frontopolar sites and O1/O2 over occipital sites & Combines frontal task-control evidence with visual and perceptual context while retaining a scalable four-electrode protocol. \\
Task battery & Twelve continuous-block, role-eligible cognitive tasks & Samples distinct constructs while minimizing movement artefact and preserving enough continuous EEG for task-level estimation. \\
Baseline comparison & Same-eye-state baselines: eyes-open tasks use an eyes-open fixation baseline; eyes-closed tasks use an eyes-closed resting baseline & Prevents the normal EEG difference between open and closed eyes from being misread as a task effect. \\
Ability evidence & Three-field contract separating theoretical, behaviourally validated, and allowed ability candidates & Requires participant performance and usable EEG evidence before an ability is credited. \\
Ability space & Twenty-three production abilities & Restricts occupational matching to cognitive abilities that the current task, behavioural, and EEG evidence can support. \\
Role bank & 376 occupations: 310 professional/knowledge-work and 66 trades/field-operations roles & Provides broad role coverage while keeping primary and supplementary pathways separate. \\
Confidence control & Session, signal, convergence, evidence, scope, and validation caps & Keeps later claims within the limits set by the available evidence. \\
Verification & 344 automated regression tests plus end-to-end session verification & Checks that the software connects the layers correctly and applies evidence rules, role scoring, and confidence caps as intended. \\
\bottomrule
\end{tabularx}
\caption{Current Neuroprofile measurement and evidence architecture.}
\label{tab:current_architecture}
\end{table}

\subsection{What the AI Components Do}

The \neuroprofile{} is not produced by one general-purpose AI model. Most decisions that affect occupational evidence are rule-based. Signal processing extracts EEG features. Evidence checks decide whether a result can be used. Fixed mappings connect qualified abilities to the role bank. Portfolio rules control which roles are shown and how confidence is limited.

AI is used in narrower places. The association agent summarizes survey themes. The implicit agent turns structured task evidence into readable descriptions. A rubric grader returns scores for the two free-text tasks. The final synthesis agent writes the user-facing report from an approved summary of the evidence. None of these AI components may create a new ability, invent a role, or override an evidence check.

\subsection{Backend Integration Safeguards}

The software checks the handoff between each layer. A blocked label applies only to the task that declares it. Confidence terms are converted to a common vocabulary before they move downstream. A participant's performance must support an ability before the EEG result can be used for that ability. The role matcher also uses the same identifiers produced earlier in the pipeline. These controls reduce the risk that one task suppresses unrelated evidence or that two layers use the same word to mean different things.

One remaining vocabulary-alignment issue between task-derived characteristics and the work-style scorer is listed in Section~\ref{sec:future_work}.

% -----------------------------------------------------------------------------
\section{Evidence Streams and Synthesis Policy}

The three streams answer different questions. Self-report describes what the participant says. Association responses describe what themes the participant is drawn towards. Task-and-EEG evidence describes what happened during scored cognitive work. They are combined in the final narrative, but they do not have the same authority in role matching.

\subsection{Rational Self-Report}

Self-report records conscious goals, preferences, and aspirations. It is useful because those choices matter in real career decisions. At the same time, it is directly affected by self-presentation and by the limits of introspection. It therefore provides context and cannot create occupational abilities or role candidates.

\subsection{Association-Based Semantic Positioning}

The association stream uses responses to images, brands, descriptors, and concepts to place the participant within the \mindspeller{} semantic network. A forward association score records the direction and strength of a response path. The route from concept A to concept B is not assumed to be the same as the reverse route.

The resulting position may indicate themes such as structure, novelty, playfulness, precision, autonomy, status, care, aesthetics, challenge, or risk sensitivity. These themes help explain motivation and preference. They do not create \onet{} abilities or roles.

\subsection{Task-and-EEG Evidence}

The task-and-EEG stream is the source of occupational evidence. It combines two kinds of observation. Behavioural scoring shows whether the participant completed the relevant demand. EEG shows the task-related neural state recorded while that demand was present.

For some abilities, such as written comprehension, speech recognition, written expression, or reaction time, the behavioural result carries the main evidential weight. EEG provides supporting context rather than proving the ability on its own.

\subsection{The 5/35/60 Synthesis Policy}

The final narrative currently gives about 5\% of its emphasis to self-report, 35\% to association evidence, and 60\% to task-and-EEG evidence. \mindspeller{} chose these values to make observed task performance and physiology the dominant source while retaining conscious and motivational context.

The percentages are not reliability estimates and are not probabilities that a statement is correct. They do not change the rules used to qualify an ability or role, and they cannot override a confidence cap. They should be revisited after empirical validation data are available.

\callout{What the weights do}{The 5/35/60 policy shapes how the final report is written. It does not decide whether a participant has earned an ability or whether a role enters consideration.}

% -----------------------------------------------------------------------------
\section{Measurement System and Feature Evidence}

\subsection{Four-Electrode EEG Recording Configuration}

The protocol records from four scalp electrodes: Fp1 and Fp2 over the frontopolar region, and O1 and O2 over the occipital region. The technical term for this electrode arrangement is an EEG \emph{montage}. Frontal channels provide information about task-related control, working-memory demand, attention, and internal processing. Occipital channels add information about visual engagement and perceptual search.

Four electrodes provide a practical, sparse recording setup. They do not support precise source localization, a full account of brain-network connectivity, direct measurement of motor-cortex activity, or comprehensive coverage of language and auditory cortex \cite{cavanagh2014,jensen2002,klimesch1999,obleser2012}.

\subsection{Continuous-Block Acquisition}

Each task contains one scored EEG block that runs without interruption. Instructions are given before the block. Speaking, typing, and button presses are delayed until after the scored period wherever the task allows. The design measures a sustained task state rather than a response-locked event-related potential. Event markers record timing, but the system does not claim ERP components tied to individual responses.

A task is accepted only when it contains enough continuous clean data. An incomplete stimulus, wrong eye condition, missing baseline, profile mismatch, or insufficient clean signal triggers a repeat instead of a weakened interpretation.

\subsection{Sampling, Filtering, and Feature Families}

EEG is sampled at 512 Hz. Preprocessing retains approximately 1--45 Hz and rejects 50 Hz line noise. The signal is analysed in two-second windows with 50\% overlap. Multitaper spectral estimation is used where available, with a conventional periodogram as fallback \cite{thomson1982,percival1993,internaldsp,internalredesign}.

For each window, the software calculates about 70 values. These include power in the delta, theta, alpha, beta, and gamma ranges; relative power; ratios between bands; entropy; total power; peak descriptors; and indicators of high-frequency artefact. Only selected feature families can influence ability evidence: spectral power, relative power, band ratios, entropy or complexity, and total-power summaries. Peak descriptors remain available for technical reporting but are not mapped directly to abilities. Gamma is withheld when the EMG check suggests contamination from facial muscle activity \cite{whitham2007,goncharova2003}.

\subsection{Same-Eye-State Baselines and Task-Evoked Change}

Open and closed eyes produce large normal differences in EEG. To avoid treating those differences as a task effect, every task is compared with a baseline recorded in the same eye condition. Eyes-closed tasks use an eyes-closed resting baseline. Eyes-open tasks use an eyes-open fixation baseline.

The main feature is the change from that personal baseline:

\[
\Delta_f^{(t)} = \mu_f^{(t)} - \mu_f^{(B)}.
\]

Here, \(\mu_f^{(t)}\) is the participant's mean value for feature \(f\) during task \(t\), and \(\mu_f^{(B)}\) is the mean for the matching baseline. The report also records normalized change, effect size, direction, block counts, and statistical uncertainty. This within-person comparison reduces the influence of stable differences in participants, device gain, and electrode contact \cite{klimesch1999,cohen1988}.

\subsection{Evidence Gating}

Before a feature can be used, it must pass a set of checks. These cover signal quality, direction of change, statistical evidence, and whether the size of the change is practically meaningful. A poor-quality result can lower the evidence strength. It cannot turn a weak result into a strong one. Multiple-comparison control and dependence-aware aggregation are used because many EEG features move together \cite{welch1947,benjamini1995,fisher1925,kost2002}.

\callout{What an EEG result can support}{The EEG analysis can show that a measured feature changed during a task relative to the participant's matching baseline. It cannot, by itself, prove a stable trait, an occupational ability, a precise brain source, or a clinical diagnosis.}

% -----------------------------------------------------------------------------
\section{O*NET Ability Measurement Scope}

\subsection{How the Ability Set Was Chosen}

\onet{} defines 52 abilities. \mindspeller{} reviewed all 52 before choosing the task battery. An ability was considered for the production set by asking five practical questions:

\begin{enumerate}
    \item Does the ability help distinguish one occupation from another?
    \item Does it matter across enough occupations or role families to be useful?
    \item Can the current tasks and sensors produce relevant evidence for it?
    \item Could the label be misunderstood as a capability that was not measured?
    \item Does it add information beyond another ability already in the set?
\end{enumerate}

This review used the O*NET content model, occupational ability data, the limits of the four-electrode protocol, and established work on human abilities and feature selection \cite{onetcontent,peterson2001,carroll1993,guyon2003,peng2005}. Exact internal weights are kept in the technical implementation record. The public point is simpler: an ability is not included just because it varies across occupations. The instrument must also be able to measure it without inviting a misleading claim.

\subsection{How Abilities Are Classified}

Each ability has one of four labels:

\begin{itemize}
    \item \textbf{EEG/task-measurable}: task performance and gated EEG can jointly support it.
    \item \textbf{Multimodal}: behavioural, linguistic, auditory, writing, or timing evidence is decisive; EEG provides task context.
    \item \textbf{Caveat-only}: the ability may matter to a role, but it can only be reported as unmeasured and never improves the match.
    \item \textbf{Blocked}: the current protocol cannot use it as positive evidence. This group mainly contains physical, psychomotor, strength, or body-control abilities.
\end{itemize}

\subsection{Production Ability Space}

The production set contains 23 abilities. Seventeen are EEG/task-measurable and six are multimodal. Oral Expression and Speech Clarity are not included because the battery does not score spoken production and the four-electrode recording cannot establish speech-production quality.

\begin{table}[H]
\centering
\small
\begin{tabularx}{\textwidth}{>{\RaggedRight\arraybackslash}p{0.25\textwidth}X}
\toprule
\textbf{Evidence class} & \textbf{Production abilities} \\
\midrule
EEG/task-measurable (17) & Inductive Reasoning; Deductive Reasoning; Fluency of Ideas; Mathematical Reasoning; Information Ordering; Originality; Category Flexibility; Visualization; Number Facility; Problem Sensitivity; Selective Attention; Memorization; Time Sharing; Spatial Orientation; Flexibility of Closure; Perceptual Speed; Speed of Closure \\
Multimodal (6) & Written Comprehension; Oral Comprehension; Written Expression; Reaction Time; Speech Recognition; Auditory Attention \\
\bottomrule
\end{tabularx}
\caption{Current 23-ability production space. Multimodal abilities require decisive behavioural, linguistic, auditory, writing, or timing evidence; EEG is contextual support rather than sole evidence.}
\label{tab:production_abilities}
\end{table}

The other 27 \onet{} abilities are blocked or caveat-only. Examples include physical strength, manual dexterity, body coordination, control precision, and low-level sensory abilities. They may appear in a caveat when they are important to a job, but they never increase a role match.

% -----------------------------------------------------------------------------
\section{Twelve-Task Production Battery}

\subsection{Design Constraints}

The twelve tasks were selected to cover the 23-ability set while staying within the limits of Fp1/Fp2/O1/O2 recording \cite{internalredesign}. Each task must:

\begin{enumerate}
    \item be compatible with the four-electrode recording;
    \item provide one uninterrupted scored block with enough clean EEG;
    \item keep speaking, typing, and movement outside the scored block where possible;
    \item add information that is not already supplied by another task;
    \item include explicit behavioural evidence when EEG context alone cannot establish the ability.
\end{enumerate}

All twelve production tasks can contribute to role matching when their behavioural and EEG checks pass. Affective appraisal, body-awareness, and colour-preference tasks may still be used as optional context modules, but they do not add occupational abilities under the current protocol.

\subsection{Task Inventory}

\begin{longtable}{>{\centering\arraybackslash}p{0.03\textwidth}>{\RaggedRight\arraybackslash}p{0.22\textwidth}>{\centering\arraybackslash}p{0.09\textwidth}>{\RaggedRight\arraybackslash}p{0.255\textwidth}>{\RaggedRight\arraybackslash}p{0.285\textwidth}}
\toprule
\textbf{\#} & \textbf{Task} & \textbf{Scored / eye state} & \textbf{Behavioural evidence} & \textbf{Candidate abilities} \\
\midrule
\endfirsthead
\toprule
\textbf{\#} & \textbf{Task} & \textbf{Scored / eye state} & \textbf{Behavioural evidence} & \textbf{Candidate abilities} \\
\midrule
\endhead
1 & Adaptive Numerical Reasoning and Sequencing & 60 s / EC & Final value, signed error, exact-answer result & Mathematical Reasoning; Number Facility; Information Ordering; Deductive Reasoning \\
2 & Working-Memory Manipulation & 60 s / EC & Normalized sequence, item-error count, exact-sequence result & Memorization; Information Ordering; Deductive Reasoning \\
3 & Auditory Target Counting & 60 s / EC & Target count, count error, exact-count result & Selective Attention; Auditory Attention \\
4 & Semantic Induction and Category Switching & 60 s / EC & Rule correctness and switch detection & Inductive Reasoning; Category Flexibility \\
5 & Visuospatial Transformation and Orientation & 60 s / EO & Grid position, distance error, orientation match & Visualization; Spatial Orientation \\
6 & Divergent Ideation & 75 s / EC & Idea count and text scored under a fixed rubric & Category Flexibility; Fluency of Ideas; Originality \\
7 & Dual-Task Performance and Rule Switching & 60 s / EC & Count error, update error, dual-task and switch cost & Time Sharing; Category Flexibility; Deductive Reasoning; Selective Attention; Information Ordering \\
8 & Rule-Based Anomaly Detection & 60 s / EO & Anomaly count, type recall, graded exact/close/miss result & Problem Sensitivity; Deductive Reasoning; Selective Attention; Information Ordering \\
9 & Rapid Visual Comparison & 60 s / EO & Rendered mismatch onset, response latency, false-alarm result & Perceptual Speed; Reaction Time \\
10 & Pattern Closure under Visual Noise & 60 s / EO & Target correctness, response latency, visibility fraction & Speed of Closure; Flexibility of Closure \\
11 & Speech-in-Noise Comprehension & up to 60 s / EC & Main idea and key detail scored against an answer key & Oral Comprehension; Speech Recognition; Auditory Attention \\
12 & Written Comprehension and Concise Synthesis & 90 s / EO & Main-idea accuracy, short written summary, rubric scores & Written Comprehension; Written Expression; Inductive Reasoning; Information Ordering \\
\bottomrule
\caption{Current twelve-task production battery. EO = eyes open; EC = eyes closed. Responses are collected after the scored EEG block wherever possible.}
\end{longtable}

\subsection{How a Task Earns an Ability: the Three-Field Contract}

Every task reports three separate ability fields:

\begin{enumerate}
    \item \textbf{Theoretical candidates}: the abilities the task was designed to examine. This describes the task design only.
    \item \textbf{Behaviourally validated candidates}: the abilities supported by the participant's scored answer, accuracy, or timing.
    \item \textbf{Allowed candidates}: behaviourally supported abilities for which the EEG is also usable and passes the relevant evidence checks.
\end{enumerate}

Only the third field enters the ability pool used for role matching. This rule prevents two participants from receiving the same ability credit when one performed the task correctly and the other did not, even if both showed an EEG change.

\subsection{Graded Behavioural Evidence}

Some results are graded rather than simply passed or failed. Speed-based tasks use latency bands. Accuracy-based tasks may distinguish exact, near, and failed responses. A weaker grade can reduce the evidence quality, but a strong grade does not create a bonus beyond a valid unpenalized result. The current bands and cut-offs are controlled pilot settings and still require empirical calibration.

\subsection{Rubric-Based Free-Text Scoring}

Tasks 6 and 12 produce free text. A language model scores the response against fixed rubric dimensions and returns the raw values. It does not decide whether the participant passes an ability. Deterministic rules make that decision after the rubric result is returned. Missing or malformed model output receives no score rather than a default pass. In this role, the model is a scoring aid, not a career recommender \cite{zheng2023}.

\subsection{Session Progression and Parallel Forms}

The session design serves two purposes. Repeating core constructs in parallel forms allows the system to examine stability without reusing the same stimuli. Adding new tasks increases ability coverage.

\begin{table}[H]
\centering
\small
\begin{tabularx}{\textwidth}{>{\centering\arraybackslash}p{0.12\textwidth}>{\RaggedRight\arraybackslash}p{0.38\textwidth}>{\centering\arraybackslash}p{0.16\textwidth}X}
\toprule
\textbf{Session} & \textbf{Content} & \textbf{Approx. protocol time} & \textbf{Cumulative ability coverage} \\
\midrule
1 & Matched baselines plus Tasks 1--4 & 6 minutes & 9 of 23 abilities \\
2 & Parallel-form repetition of Session 1 plus Tasks 5--9 & 15 minutes & 17 of 23 abilities \\
3 & Parallel-form repetition of Sessions 1--2 plus Tasks 10--12 & 21 minutes & 23 of 23 abilities \\
\bottomrule
\end{tabularx}
\caption{The session design trades off repeated evidence and broader construct coverage.}
\end{table}

Completing more sessions does not automatically raise confidence. It increases the amount of evidence that could support a stronger conclusion. Confidence rises only when the repeated behaviour, EEG quality, cross-task support, and validation rules also justify it.

% -----------------------------------------------------------------------------
\section{O*NET Role Mapping and Role Bank}

\subsection{Why O*NET Is Used}

\onet{} provides a common vocabulary for occupational abilities and reports how important each ability is across 894 occupations. \mindspeller{} uses release 30.3 of the database so that the occupational reference can be reproduced later. Role matching considers only requirements that fall within the measurable 23-ability set \cite{onetdb,onetcontent,peterson2001,fleishman1992}.

\subsection{How Roles Enter the Bank}

An occupation does not enter the bank merely because it shares one ability with the participant. During curation, \mindspeller{} considers how many of the role's important cognitive requirements are measurable, how strongly those requirements are represented, whether several abilities add independent information, and whether the remaining unmeasured requirements could lead readers to overstate what the system has shown.

Professional and knowledge-work occupations must have broad coverage within the measurable ability set. Trades and field-operations occupations can enter with narrower cognitive coverage because the claim is narrower: the report describes cognitive overlap only, not practical readiness for the trade or field role.

The exact coefficients and cut-offs remain in the internal technical record. The public commitment is that the same documented rules are applied to every occupation and can remove a role as well as include it.

\subsection{Role-Bank Composition}

The internal role bank contains \textbf{376 occupations} \cite{internalredesign}:

\begin{table}[H]
\centering
\small
\begin{tabular}{lrr}
\toprule
\textbf{Role-bank category} & \textbf{Occupations} & \textbf{Interpretive use} \\
\midrule
Professional / knowledge-work & 310 & Primary cognitive-fit role pathways \\
Trades / field operations & 66 & Supplementary cognitive-fit or mindset pathways \\
\midrule
Total & 376 & Internal candidate bank before portfolio selection \\
\bottomrule
\end{tabular}
\caption{Current internal role-bank composition.}
\label{tab:role_bank_composition}
\end{table}

The 23 abilities cover analytical, technical, language-based, perceptual, inspection, operations, and field-role families. A role enters the bank only when its measurable cognitive requirements meet the rules for its bank. Requirements outside the protocol remain visible as caveats.

\subsection{Eligibility, Selection, and Role Confidence}

The role layer answers three separate questions:

\begin{itemize}
    \item \textbf{Eligibility}: does the participant have enough qualified ability evidence for this role to be considered?
    \item \textbf{Selection}: should the role appear in the short user-facing portfolio after ranking, overlap, and portfolio balance are considered?
    \item \textbf{Role confidence}: how strongly may the report describe the cognitive overlap after all caps are applied?
\end{itemize}

A role can be eligible and still be left out of the final portfolio when several occupations express essentially the same pattern. An adjacent pathway has some relevant evidence but does not meet the stronger standard used for selected roles. The exact thresholds are controlled implementation settings and are not published in this white paper.

\subsection{Portfolio Policy and Supplementary Pathways}

The final \neuroprofile{} does not list every eligible occupation. It selects a small portfolio in which professional and knowledge-work roles form the clear majority. Trades and field-operations roles are limited to a minority share, with an approximate 80/20 portfolio target.

In this document, \textbf{supplementary} describes the purpose of a pathway, not a weaker job recommendation. A supplementary pathway is a mindset illustration. It shows a type of work that uses a similar measured pattern. For example, a Power Plant Operator pathway may illustrate sustained monitoring, information ordering, and divided attention. It does not state that the participant knows plant systems, can act safely in an emergency, holds the required licence, or is ready to perform the job. Those requirements remain caveats and never raise the match.

\callout{How to read a supplementary pathway}{Read it as ``this work setting uses a similar cognitive pattern,'' not as ``this person should take this job.''}

\subsection{Role Matching Is Not Hiring Suitability}

The role layer compares cognitive requirements only. It does not combine experience, qualifications, work samples, references, communication, compensation, legal eligibility, or organizational need into a hire/no-hire result. The \neuroprofile{} can support a recruitment conversation, but it does not decide whether someone should be hired.

% -----------------------------------------------------------------------------
\section{Work-Style Interpretation}

Many occupations draw on the same small group of abilities. To avoid repeating a long list of similar job titles, the Work-Style Interpretation Layer summarizes the recurring pattern in plain work terms.

Examples include:

\begin{itemize}
    \item \textbf{Structured reasoning and sequencing}: Information Ordering, Deductive Reasoning, and working-memory evidence.
    \item \textbf{Monitoring and anomaly detection}: Selective Attention, Auditory Attention, Problem Sensitivity, and task consistency.
    \item \textbf{Perceptual comparison and closure}: Perceptual Speed, Speed of Closure, and Flexibility of Closure.
    \item \textbf{Spatial transformation}: Visualization and Spatial Orientation.
    \item \textbf{Comprehension and concise synthesis}: Oral or Written Comprehension, Written Expression, Inductive Reasoning, and Information Ordering.
    \item \textbf{Workflow switching}: Time Sharing and Category Flexibility.
    \item \textbf{Controlled ideation}: Fluency of Ideas, Originality, and Category Flexibility.
    \item \textbf{Quantitative handling}: Mathematical Reasoning and Number Facility.
\end{itemize}

These themes summarize measured evidence; they are not new performance measures. A monitoring theme does not prove vigilance in a safety-critical workplace. A delivery-readiness proxy does not measure delivery speed, and a diligence-related proxy does not establish work ethic.

% -----------------------------------------------------------------------------
\section{Confidence, Caps, and Current Validation Status}
\label{sec:confidence}

\subsection{Four Different Confidence Concepts}

The document keeps four questions separate:

\begin{itemize}
    \item \textbf{Signal reliability}: is the EEG recording, including its baseline, technically usable?
    \item \textbf{Ability evidence quality}: does a specific ability have valid behavioural and EEG support?
    \item \textbf{Role confidence}: how strongly can the report describe the overlap with one occupation?
    \item \textbf{Neuroprofile Match Reliability}: how coherent is the final interpretation across the evidence streams that were available?
\end{itemize}

A large EEG change does not automatically produce a strong ability claim. A coherent \neuroprofile{} does not automatically make every role high confidence. These are different judgements made at different points in the process.

\subsection{How Conservative Caps Work}

A confidence cap sets the highest level that a later claim may reach. The system uses the lowest applicable limit:

\[
C_{\mathrm{role}} \leq \min\bigl(C_{\mathrm{session}}, C_{\mathrm{signal}}, C_{\mathrm{convergence}}, C_{\mathrm{evidence}}, C_{\mathrm{scope}}, C_{\mathrm{validation}}\bigr).
\]

In plain language, good performance in one part of the system cannot cancel a serious limitation elsewhere. A clean Session 3 recording may still be limited by pilot validation status. Poor signal quality may lower the result further. The formula states this rule without disclosing the proprietary scoring coefficients.

\subsection{Where Caps Apply}

Caps apply when task results are turned into characteristics, when characteristics are turned into ability evidence, when roles are described, and when the final occupational text is written. They do not alter the raw feature statistics. A feature may show a strong task-related change while the resulting role statement remains Moderate or Low confidence.

The final synthesis agent must reproduce the cap it receives. It cannot promote an adjacent role, remove a caveat, or add evidence that was not present upstream.

\subsection{The Current Pilot Cap}

The current protocol is marked \texttt{validation\_status: pilot}. Role confidence is therefore capped at \textbf{Moderate for every participant}, even after a complete session with technically clean data. This is a product safeguard, not a statement that all participants have the same data quality. High role confidence can become available only after the protocol has been calibrated on a participant sample and its validation status has been formally changed.

\subsection{What Is Already Implemented and Verified}

The software already checks signal quality, matching baselines, likely EMG contamination, task-to-baseline changes, effect sizes, uncertainty, multiple comparisons, dependent feature evidence, behavioural validity, task-specific blocked labels, and role caveats.

The backend also has 344 automated tests covering report parsing, ability mapping, role matching, rubric grading, characteristic scoring, weighted abilities, and downstream screening. One end-to-end run using a real Session 3 recording produced 22 of the 23 abilities, kept the relevant blocked labels in place, scored all 376 role candidates, created the compact portfolio, and respected the Moderate cap.

This is \textbf{functional verification}: it shows that the implemented parts connect and behave as specified. It is not evidence that the resulting scores are psychometrically valid or predict work outcomes.

\subsection{What Has Not Yet Been Established}

The following questions still require empirical study:

\begin{itemize}
    \item Do participants obtain stable results when they repeat parallel versions of the tasks?
    \item Do the tasks agree with established measures of the same cognitive constructs?
    \item Are the current accuracy, speed, and rubric cut-offs appropriate for the intended population?
    \item Do independent human raters agree with the free-text rubric grader?
    \item Is the speech-in-noise task delivered consistently across supported devices?
    \item Do independent experts agree on the ability measurement labels and claim-risk ratings?
    \item Do the results add useful information about training, performance, retention, or hiring outcomes beyond existing methods?
\end{itemize}

Until those studies are complete, \mindspeller{} does not claim validated prediction of job performance, validated hiring suitability, or superiority over established selection systems.

\subsection{How Confidence Can Increase}

Confidence can rise only when the evidence improves. Relevant steps include calibrating thresholds on participant data, testing repeated-session reliability, comparing tasks with established instruments, checking agreement between human raters and the rubric grader, standardizing deployment conditions, and approving a revised protocol profile.

A later outcome study may test whether task-and-EEG evidence adds information beyond self-report and association evidence. No such result is claimed in this white paper.

\callout{Verification versus validation}{Software tests can show that the pipeline applies its rules correctly. They cannot show that the final result predicts success at work. That requires external validation data.}

% -----------------------------------------------------------------------------
\section{AI Components, Synthesis, and Guardrails}

\subsection{Where AI Is Used}

\begin{table}[H]
\centering
\small
\begin{tabularx}{\textwidth}{>{\RaggedRight\arraybackslash}p{0.24\textwidth}>{\RaggedRight\arraybackslash}p{0.34\textwidth}X}
\toprule
\textbf{Component} & \textbf{Permitted function} & \textbf{Not permitted} \\
\midrule
Association / explicit agent & Summarize association-derived preferences, values, and semantic position & Generate abilities or role candidates \\
Implicit interpretation agent & Translate structured task evidence into task-supported characteristics & Invent abilities, ignore behavioural failure, or reinterpret blocked labels \\
Free-text rubric grader & Return raw rubric scores for Tasks 6 and 12 under fixed dimensions & Decide pass/fail, set thresholds, or see unrelated answers \\
Role matching layer & Deterministically compare the qualified ability pool with the curated role bank & Free-form career generation \\
Final synthesis agent & Turn the approved evidence summary into a readable \neuroprofile{} & Rename or invent roles, raise confidence, remove caveats, or make job-performance claims \\
\bottomrule
\end{tabularx}
\caption{AI helps interpret and write the report; ability and role qualification remain rule-based.}
\end{table}

The table separates language-model tasks from rule-based decisions. In particular, the role matching layer is deterministic. The final synthesis agent receives role results; it does not create them.

\subsection{Compact Handoff}

The final synthesis agent does not receive raw EEG arrays or all 376 role records. It receives a compact, structured handoff containing the information it is allowed to discuss: self-report context, association themes, task-supported characteristics, qualified abilities, selected and adjacent pathways, work-style themes, confidence caps, caveats, and blocked claims.

This design reduces token use and makes the output easier to audit. A reviewer can compare the handoff with the final text and check whether the model stayed within the supplied evidence.

\subsection{Final Neuroprofile Output}

The final \neuroprofile{} may include a combined cognitive and motivational summary, decision-making style, task-evidence themes, comparison between reported and measured patterns, career domains, preferred work environments, selected and supplementary pathways, development suggestions, confidence statements, and data structures for frontend charts.

These sections are different views of the same handoff. They are not separate analyses, and repetition across them should not be treated as additional evidence.

% -----------------------------------------------------------------------------
\section{Customization and Responsible Expansion}
\label{sec:customization}

\subsection{The Battery as a Reference Protocol}

The twelve-task battery is a reference protocol, not a permanently fixed test form. Employers and research partners may use parallel versions, adjust difficulty, choose session depth, or propose new modules. Each change must preserve the link between the task, the construct it measures, the participant's behavioural result, the EEG context, and the ability that may be reported.

Permitted changes include:

\begin{itemize}
    \item changing stimuli while keeping the same construct and scoring rules;
    \item adjusting difficulty for the target population without using difficulty as a substitute for seniority;
    \item using parallel forms to reduce memory effects across sessions;
    \item adding role-relevant semantic, visual, auditory, or rule-based material;
    \item adding a new task only after its target ability, behavioural score, EEG role, blocked claims, and reliability requirements have been specified.
\end{itemize}

\subsection{Compatibility with the Four-Electrode Recording Configuration}

New modules must remain suitable for Fp1/Fp2/O1/O2 recording. The strongest candidates use a continuous, low-movement task state, standardized stimulus delivery, a baseline with the same eye condition, and delayed responses. Tasks dominated by continuous speech, uncontrolled movement, fine motor execution, or neural systems not covered by the four electrodes need other sensors or should remain behavioural-only.

\subsection{Role-Bank Expansion Follows Ability Evidence}

Role coverage should grow in this order:

\begin{center}
ability needed \pipelinearrow{} task designed and tested \pipelinearrow{} evidence rules defined \pipelinearrow{} ability allowed \pipelinearrow{} role bank updated.
\end{center}

A role should not be added first and justified afterward. When its relevant abilities are already supported, inclusion is mainly a role-bank calibration and coverage decision. When a central ability is not supported, the task battery must be expanded before that ability can influence role matching.

\subsection{Remaining Scope Gaps}

Oral Expression and Speech Clarity are not part of the current ability set because the battery does not score spoken production and the four-electrode recording cannot establish speech-production quality. Social Perceptiveness, Persuasion, Service Orientation, and related interpersonal constructs also remain blocked because there is no validated social-cognition task.

Physical strength, manual skill, body control, and low-level sensory abilities remain outside the cognitive-fit protocol. They are not candidates for inference from sparse EEG. Optional affective appraisal, body-awareness, and colour-preference modules may add context, but they cannot raise an occupational ability score without a validated evidence route.

% -----------------------------------------------------------------------------
\section{Future Work}
\label{sec:future_work}

The next phase should focus on calibration and validation rather than simply increasing the number of roles.

\begin{enumerate}
    \item \textbf{Calibrate pilot thresholds.} Replace constructed accuracy, speed, and rubric cut-offs with values based on participant distributions.
    \item \textbf{Test repeated-session reliability.} Measure how stable behavioural results and EEG feature changes are when participants complete parallel forms \cite{shrout1979}.
    \item \textbf{Check construct agreement.} Compare the production tasks with established measures of mental rotation, perceptual speed, closure, sustained attention, divergent thinking, working memory, and speech-in-noise comprehension.
    \item \textbf{Measure rubric agreement.} Compare the free-text grader with independent human raters before rubric-dependent abilities are used beyond exploratory reporting.
    \item \textbf{Repeat the ability review.} Use an independent rater for measurement labels and claim risk, and refine the redundancy analysis using occupation-by-ability data.
    \item \textbf{Standardize speech-in-noise delivery.} Verify acoustic level and timing across supported hardware.
    \item \textbf{Complete vocabulary alignment.} Finish the remaining mapping between task-derived characteristic identifiers and the HR/work-style scorer.
    \item \textbf{Study external validity.} After the measurement model is stable, test whether the results relate to appropriate training or work outcomes and whether they add information beyond existing methods. Internal role matches and \onet{} ratings should not be treated as ground truth \cite{aera2014,messick1995,schmidt1998}.
\end{enumerate}

The 5/35/60 narrative policy should be tested in the same programme. Until then, it remains a \mindspeller{} product setting rather than a validated optimum.

% -----------------------------------------------------------------------------
\section{Glossary}

\begin{longtable}{>{\RaggedRight\arraybackslash}p{0.285\textwidth}>{\RaggedRight\arraybackslash}p{0.64\textwidth}}
\toprule
\textbf{Term} & \textbf{Plain-language meaning} \\
\midrule
\endfirsthead
\toprule
\textbf{Term} & \textbf{Plain-language meaning} \\
\midrule
\endhead
Adjacent pathway & A role with some relevant evidence that does not meet the stronger standard used for selected roles. The exact cut-offs are internal. \\
Allowed ability candidate & An ability supported by the participant's task result and by usable EEG evidence that has passed the required checks. \\
Allowed ability pool & The set of allowed abilities from the current session. This is the only ability set used by the role matcher. \\
Association stream & Responses to images, brands, descriptors, concepts, and IAT-style prompts, used to place a participant within the \mindspeller{} semantic network. \\
Baseline-referenced delta & The difference between an EEG feature during a task and the same feature during the participant's matching baseline. \\
Behaviourally validated ability & An ability supported by the participant's scored task performance before EEG evidence is considered. \\
Blocked ability & An ability that the current protocol cannot use as positive evidence. \\
Blocked inference & A label that a specific task is not allowed to produce. A block applies only to the task that declares it. \\
Caveat & A limitation shown with a result, such as an unmeasured licence, trade skill, or sensory requirement. A caveat never improves a match. \\
Caveat-only ability & A real job requirement that may be reported as unmeasured but cannot be used as positive matching evidence. \\
Claim risk & The risk that a label could be mistaken for a capability that was not actually measured. \\
Cognitive fit & Overlap between the participant's measured cognitive abilities and the cognitive requirements of a role. It does not mean practical readiness or hiring suitability. \\
Compact handoff & The structured summary sent to the final synthesis model. It contains only the evidence, roles, caps, caveats, and blocked claims the model may discuss. \\
Confidence cap & The highest confidence level a later claim may reach. Later evidence may lower it, but the synthesis agent cannot raise it. \\
Continuous-block acquisition & Recording one sustained scored EEG interval for a task instead of analysing separate response-locked trials. \\
Deterministic layer & A processing step controlled by fixed rules rather than free-form language generation. \\
EEG recording configuration (montage) & The placement of EEG electrodes. The current configuration uses Fp1/Fp2 at frontopolar sites and O1/O2 at occipital sites. \\
EEG/task-measurable ability & An ability that can be supported by task performance together with gated EEG under the current protocol. \\
Evidence gate & The behavioural, signal-quality, direction, statistical, and change-size checks that a result must pass before it can be used. \\
Evidence stream & A parallel source of information. The three streams are self-report, association evidence, and task-and-EEG evidence. \\
Explicit--implicit alignment & The relationship between self-reported or association-derived themes and task-supported evidence: aligned, partly aligned, aspirational, or in tension. \\
Feature family & A group of related EEG measurements, such as spectral power, relative power, band ratios, or entropy and complexity. \\
Forward association score & A directional score showing how strongly a response points towards a concept in the semantic network. \\
Functional verification & Evidence that the software runs from start to finish and applies its rules correctly. It is not psychometric validation. \\
Global EEG reliability & A summary of recording quality, baseline quality, usable data, and artefact burden. It can limit later claims. \\
Hiring suitability & The full employment judgement, including experience, qualifications, work samples, communication, references, legal factors, and organizational needs. The \neuroprofile{} does not determine it. \\
Match Reliability & The overall coherence of the final \neuroprofile{} across the evidence streams available for that participant. It is separate from EEG signal reliability and role confidence. \\
Measurement label & One of four labels applied to an ability: EEG/task-measurable, multimodal, caveat-only, or blocked. \\
Measurement scope & The claims the current hardware, tasks, behavioural evidence, and validation status can support. \\
Multimodal ability & An ability for which behavioural, language, auditory, writing, or timing evidence is decisive and EEG provides task context. \\
Neuro-Anchor & The physiological contribution of the task-and-EEG stream. It adds task-related neural evidence but is not a personality or career classifier by itself. \\
Neuroprofile & The final report combining self-report, association evidence, task-and-EEG characteristics, abilities, roles, work-style themes, confidence, and caveats. \\
Occupational evidence & Evidence allowed to influence \onet{} abilities and role matching. In the current system it comes only from the task-and-EEG stream. \\
O*NET database release 30.3 & The version of the \onet{} occupational database used for the current role bank. Naming the release makes the reference reproducible. \\
O*NET Importance rating & An \onet{} rating of how important an ability is to an occupation. \\
Parallel form & A different stimulus version of the same task, used during repetition to reduce simple memory effects. \\
Pilot profile & A protocol whose stimuli, rubrics, and thresholds are still awaiting empirical calibration. \\
Processing layer & A sequential step such as feature extraction, ability mapping, role matching, or final synthesis. \\
Professional/knowledge-work bank & The role-bank category used mainly for primary cognitive-fit pathways. It currently contains 310 occupations. \\
Qualified ability & An allowed ability that also passes the downstream evidence-quality checks. \\
Role bank & The curated set of occupations scored by the role matcher. It currently contains 376 occupations. \\
Role confidence & The strongest wording permitted for one role after all applicable caps are considered. \\
Role portfolio & The short set of roles shown in the final report after ranking, overlap control, and bank-balance rules. \\
Rubric grader & A language model that returns raw scores for free-text responses under fixed criteria. Deterministic rules decide whether those scores support an ability. \\
Same-eye-state baseline & A baseline recorded with the same eye condition as the task: eyes-open fixation for an eyes-open task, or eyes-closed rest for an eyes-closed task. \\
Selection status & Whether an eligible role is shown in the final portfolio or withheld. \\
Semantic network & The graph of concepts used to place participants from their directional association responses. \\
Signal integrity & Checks for artefact, flatline data, electrode contact, wrong eye state, and other recording problems. \\
Signal reliability & The degree to which the EEG and baseline data are usable for the intended claim. \\
Supplementary pathway & A trades or field-operations role used mainly to illustrate a cognitive style. It is not a literal job recommendation and carries explicit caveats. \\
Synthesis weight & A \mindspeller{}-defined share of emphasis in the final narrative. Current values are 5\% self-report, 35\% association evidence, and 60\% task-and-EEG evidence. They are not validation coefficients. \\
Task convergence & Support for the same ability or interpretation from more than one task or characteristic family. \\
Task-supported characteristic & A plain description of what the participant's valid task performance and EEG jointly support. \\
Theoretical ability candidate & An ability a task was designed to examine. It is not credited until behaviour and EEG also support it. \\
Three-field ability contract & The separation between theoretical candidates, behaviourally supported candidates, and allowed candidates. \\
Trades/field-operations bank & The role-bank category used for supplementary cognitive-fit pathways. It currently contains 66 occupations. \\
Work-style proxy & A work-related theme summarized from repeated ability evidence. It is not a direct measure of workplace performance. \\
\bottomrule
\end{longtable}

% -----------------------------------------------------------------------------
\section{Conclusion}

The \mindspeller{} \neuroprofile{} combines three sources that answer different questions. Self-report records conscious goals and preferences. Association responses show motivational and semantic themes. Scored tasks and EEG provide the only evidence used to qualify \onet{} abilities and occupational matches.

The current method uses four electrodes, twelve tasks, a three-stage ability check, 23 production abilities, and a bank of 376 occupations. A role appears only after the participant's task performance and EEG evidence meet the relevant rules. Work-style themes summarize recurring patterns, and supplementary pathways illustrate similar thinking demands without claiming practical job readiness.

The protocol remains a pilot. Signal checks, behavioural gates, software tests, and end-to-end verification are already in place, but external reliability and validity studies are still required. Role confidence therefore remains capped at Moderate. The next priority is to calibrate the measurement system on participant data and test whether the results are stable, agree with established measures, and add useful information in real recruitment settings.

The intended use is straightforward: support a clearer conversation about measured cognitive fit while keeping qualifications, practical skills, and hiring decisions with the employer and the candidate.

% -----------------------------------------------------------------------------

\end{document}